\documentclass[%
 reprint,
 amsmath,amssymb,
 aps,
]{revtex4-2}
\usepackage{cleveref}
\usepackage[caption=false]{subfig}
\usepackage{graphicx}
\usepackage{multirow}
\usepackage{dcolumn}
\usepackage{bm}

\begin{document}

\preprint{APS/123-QED}

\title{Income Inequalities Increase with City Size:\\ Evidence from French Data}

\author{Nirbhay Patil}
 \affiliation{Laboratoire de Physique de l’\'Ecole normale supérieure, ENS, Université PSL, CNRS, Sorbonne Université, Université Paris Cité, 75005 Paris, France:\\Chair of EconophysiX and Complex Systems, Ecole Polytechnique, 
Palaiseau, France}
\author{Jean-Pierre Nadal}
 \affiliation{Laboratoire de Physique de l’\'Ecole normale supérieure, ENS, Université PSL, CNRS, Sorbonne Université, Université Paris Cité, 75005 Paris, France;\\
 Centre d'Analyse et de Mathématique Sociales, EHESS, CNRS, 75006 Paris, France}
\author{Jean-Philippe Bouchaud}%
 
\affiliation{Capital Fund Management, Paris, France; \\
Chair of EconophysiX and Complex Systems, Ecole Polytechnique, 91128 
Palaiseau, France; \\
\& Acad\'emie des Sciences, Paris, France
}%

\date{\today}

\begin{abstract}
We analyse the income distributions of cities in France and the scaling of the income of different deciles as a function of the population. We find a significant difference in the scaling exponents for the richer and poorer parts of the population, implying an unequivocal rise in inequalities in larger cities, made worse by living costs that are disproportionately higher for the poor. We find that the distribution of revenues of cities in France has a universal, Gumbel-like form, with mean and variance growing with the logarithm of population. We show how this result directly implies different income scaling exponents as a function of decile. We also study the spatial correlations of income and population, which decay exponentially with distance. We find that large cities are not more income-segregated than small cities. Finally, we search for couplings between social and economic factors, like age and income, and propose a toy model that reproduces some of our observations.
\end{abstract}

\maketitle


\section{\label{sec:level1}Introduction}

Urbanization is one of the most significant global trends of our time. Cities are engines of economic growth and development, attracting millions of additional people every year. Characterized by their size, density, and complexity, cities are home to large and diverse populations. As cities continue to grow, their complexity and the challenges they face continue to increase, as it brings with it a host of social, economic, and environmental challenges.

One of the most significant challenges of urbanization is inequality. While cities offer opportunities for economic and social mobility, they also host significant disparities in wealth and income. These inequalities can have severe consequences, impacting health, education, and quality of life, among other things. Understanding the patterns and drivers of inequality in cities is thus critical for creating sustainable and equitable urban environments, and may also shed light on the mechanisms leading to wealth and income inequalities in general. 

Many urban features, including economic, social, and environmental factors, scale with the size of population. This scaling is known as {\it urban scaling}, and it has been observed across a wide range of urban phenomena, from innovation to crime, across the world \cite{west2018scale,bettencourt2007growth,bettencourt2016urban,meirelles2018evolution,zund2019growth,sahasranaman2019urban}. However, little is known about how inequality scales with population, or how it changes as cities grow. Does the distribution of wealth or income change with the size of a city? Is inequality greater in larger cities, or do larger cities offer greater opportunities for economic mobility, hence alleviating poverty? These questions are critical for understanding the dynamics of urban inequality and developing policies to address it.

In a very interesting recent paper, Mora et al. \cite{mora2021scaling} have investigated the scaling of inequality using data from US cities. One of their main findings is that the top income deciles increases super-linearly with city size when the bottom income decile only increases sub-linearly with city size. In other words, income inequalities is markedly higher in large cities than in small cities. In the present paper, we investigate the same question using data provided by the French institute of statistics (INSEE). Our quantitative results are, perhaps surprisingly, very close to those obtained in US cities. Additional results, not addressed in Ref. \cite{mora2021scaling}, are also reported. We show in particular that the distribution of income in different cities appears to be {\it universal}, when centred to account for different means and rescaled to account for different standard deviations. We show that the dependence of decile income with city size is entirely explained by the fact that the income mean and income standard deviation both increase logarithmically with city size, albeit at different rates. The mean grows slower than the standard deviation, leading to poorer poors and richer affluents in large cities. This is in agreement with the result of Ref. \cite{cottineau2019defining}, which reports a Gini coefficient that increases as the logarithm of city sizes. We also study the spatial correlations of income and population, as well as other interesting statistical trends that differentiate small and large cities. In particular, we find that large cities are not more income-segregated than small cities, confirming the conclusion reached in \cite{cottineau2019defining}. Finally, we will discuss possible reasons for the observed scaling of inequalities with city size, and propose a toy-model that qualitatively reproduces the growth of income inequalities with city size.  

\section{Data analysis}

\subsection{Presentation of the data}

For the purpose of our analysis, we use data provided by the INSEE (Institut National de la Statistique et des Etudes Economiques). This data is geolocalised, and split into averages over boxes of side 200 meters. Thus we have the locally averaged yearly incomes across France by location in tandem with local population, number of families, living area, home ownership, and lots of other information \cite{DonCar}. The data is obtained from the tax declarations for 2015 (the forms are filed individually the year after, thus in 2016).
\begin{figure}
    \centering
    \hspace{-0.4cm}\includegraphics[width=\linewidth]{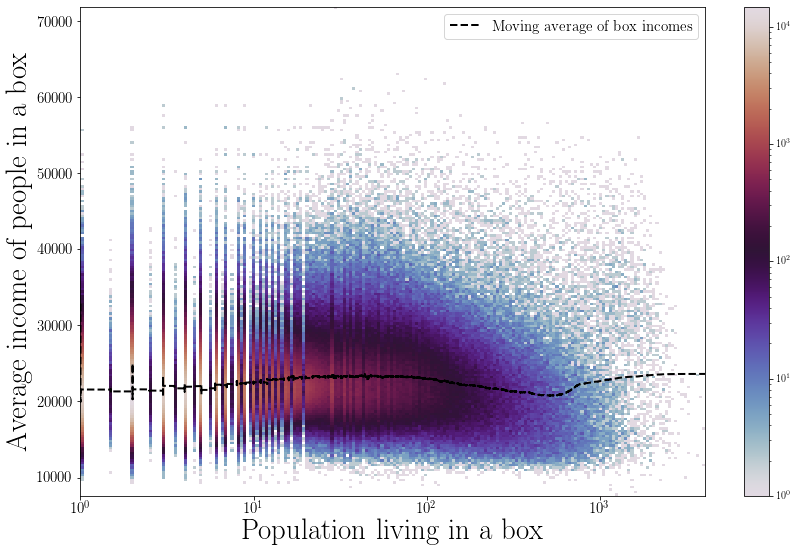}
    \caption{Average yearly standard of living in euros of each box of side 200 meters plotted against its population. The colourbar represents the density of points on the plot, and the black line is a moving average over the incomes of boxes arranged in order of increasing population.}
    \label{fig:nullcorrelationplots}
\end{figure}
\begin{figure*}[t]
  \subfloat[]{\label{fig:denscmap}%
  \includegraphics[height=6.5cm]{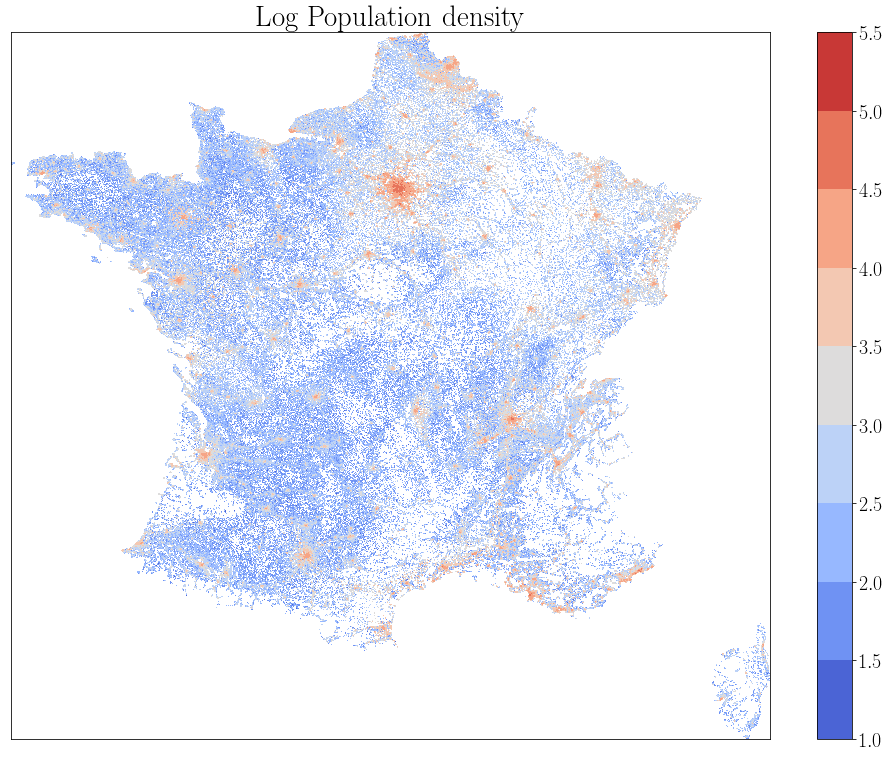}%
}
\subfloat[]{\label{fig:qolcmap}%
  \includegraphics[height=6.5cm]{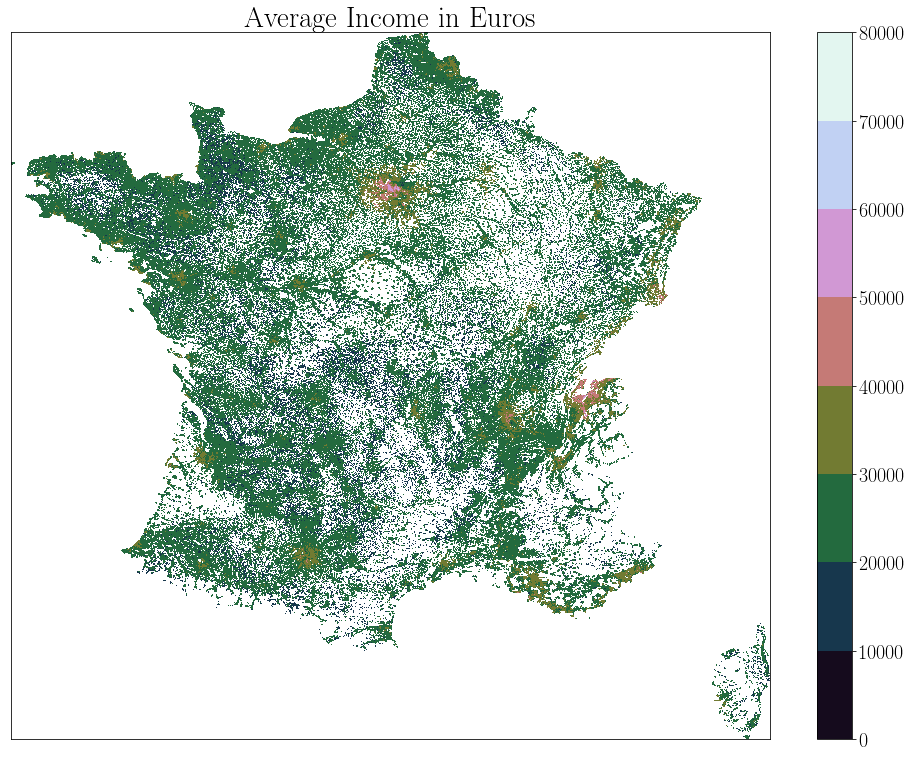}%
}
    \caption{Visualisation of the data used in this study, with the log population density in \textbf{(a)} and the average quality of life in \textbf{(b)}. Each data point represents an average over a box of side 200m located at the given location.}
\end{figure*}

It is important to note that the ``income" data we represent is in fact the ``yearly standard of living" defined by INSEE, which factors in the number of people living in a household.  In particular, a family of two people each earning the same salary as a person comprising a household alone will be shown to have a higher standard of living in euros than the latter. This is achieved by dividing the total income of each family by the number of ``units of consumption"  (UC) in the family. The first adult in the family has a UC of 1, while the rest have a UC of 0.5. Children under the age of 14 are taken to have a UC of 0.3. This standard of living winsor-summed over the whole population of each box represents a single data point, shown as a scatter plot as a function of population density in Fig. \ref{fig:nullcorrelationplots}. At first glance, we find that there is very little relation between local population density and average income (see the moving average line through the cloud of points). Though the most populated boxes are poorer, this is not a very striking observation. However, as we show now, grouping together boxes belonging to the same city leads to a much clearer picture (see also section \ref{sec:scaling-bis}

\subsection{Scaling of different income deciles with city size}\label{sec:scaling}

To see how the incomes of various sections of the population changes with city size, we apply a similar methodology as proposed in \cite{mora2021scaling}, comparing how different income deciles scale with population.


Following \cite{mora2021scaling}, the variable whose scaling we are looking for is 
\begin{align}
    Y_i^{(n)}=\sum_{j\in \mathcal{D}_n}y_{i,j}
\end{align}
where $\mathcal{D}_n$ represents the $n^{ \mathrm{th}}$ decile, and $y_{i,j}$ is the income of a person in city $i$. We are looking for scaling of the form:
\begin{align}
    Y_i^{(n)}\propto N_i^{\beta^{(n)}}.
\end{align}
When the distribution of income is independent of city size, the total income of each decile should grow proportionally to $N_i$, i.e. $\beta^{(n)} \equiv 1$. Different scaling coefficients for different deciles would characterise the dependency of inequalities on city size. 

To obtain these exponents, we perform a linear least box regression of $\log(Y_i^{(n)})=\beta^{(n)}\log(N_i)+c$. One challenge to this approach however is defining what a ``city'' is. We try various methods of clustering points and categorising them as parts of different cities, and the exponents obtained are slightly different in each case. 
\begin{figure*}
 \hspace{-0.6cm}
    \subfloat[width=0.45\linewidth][]{\begin{tabular}[b]{c  c }%
           \includegraphics[width=0.18\linewidth]{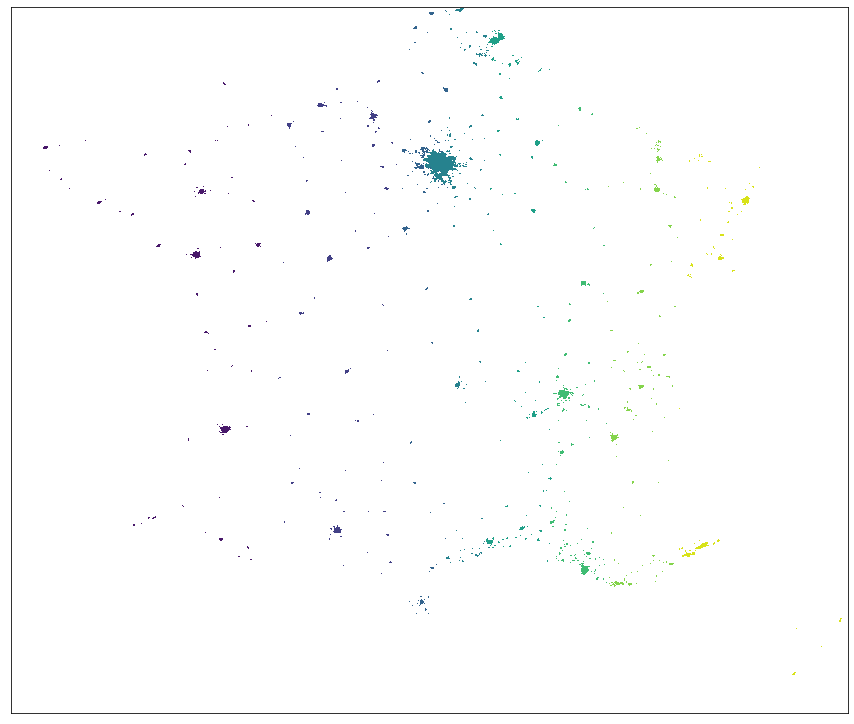} & \hspace{-0.5cm}\includegraphics[width=0.18\linewidth]{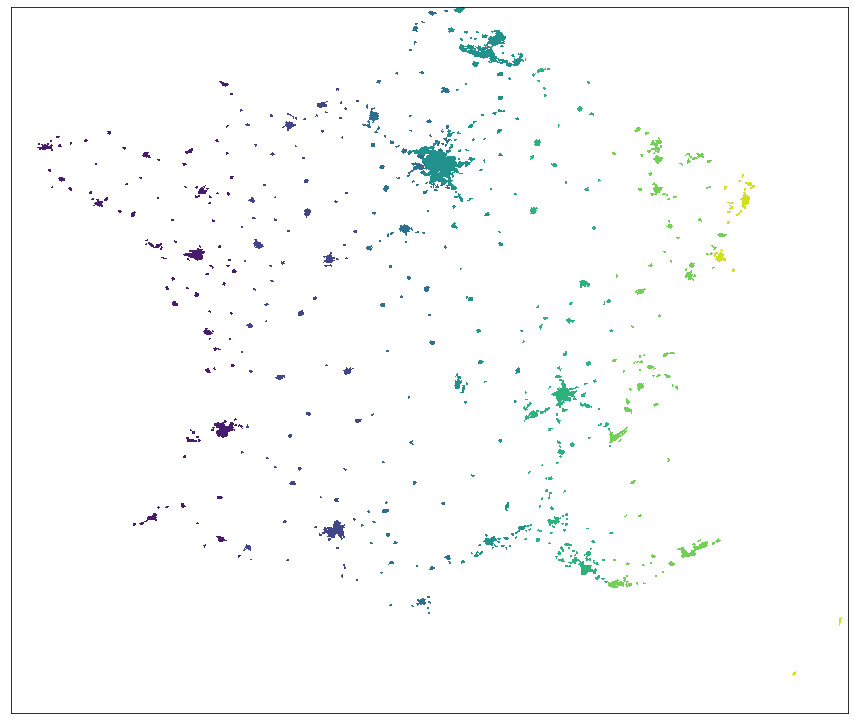}\\ 
        \includegraphics[width=0.18\linewidth]{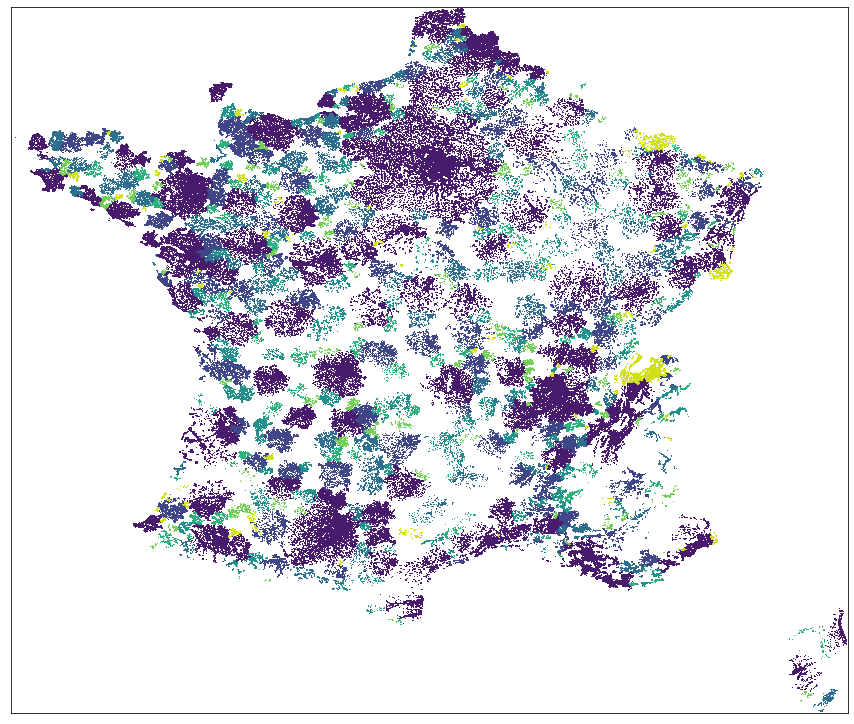} &\hspace{-0.5cm}
    \includegraphics[width=0.18\linewidth]{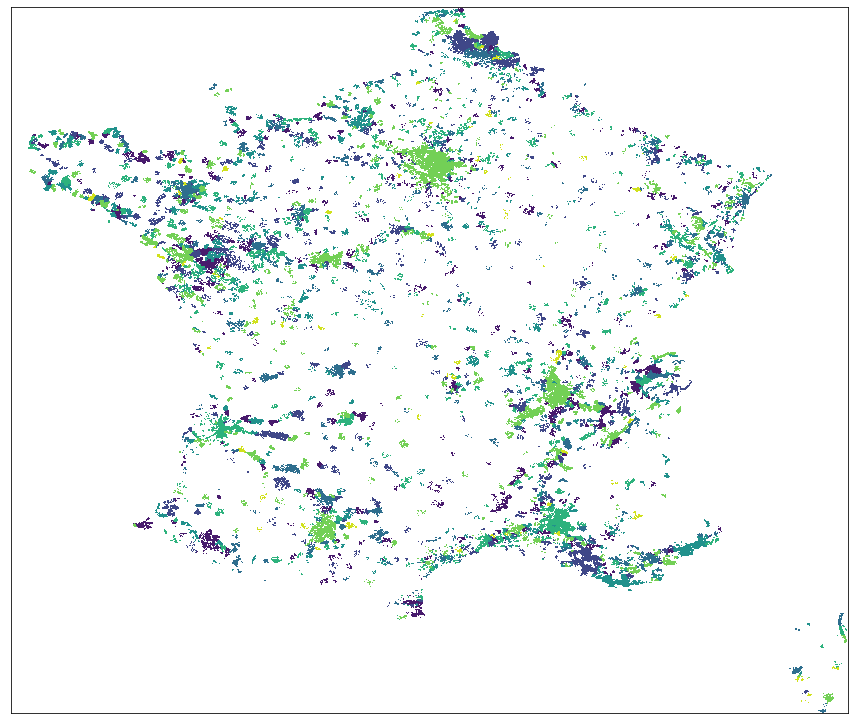}

                    \end{tabular}}
    \subfloat[width=0.3\linewidth][]{
        \centering\vspace{0.2cm}
        \includegraphics[height=6cm]{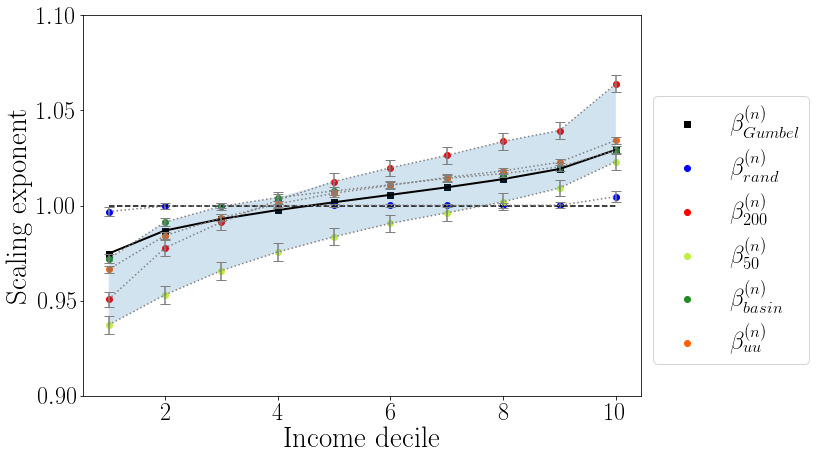}
        \centering
        }
    \caption{Scaling exponents $\beta^{(n)}$ of the incomes of different quantiles of the population \textbf{(b)} based on which definition for cities \textbf{(a)} is chosen. The top two images in \textbf{(a)} are manually performed clustering of semi-continuous regions formed of boxes with at least 200 and at least 50 people each. The bottom two images in \textbf{(a)} are two different definitions for urban regions we found on INSEE. On the left we have what is called the basins of the cities, which includes all communes that send at least 15\% of their working population to the most populated part of the basin. On the right we have ``Urban Units'', which are defined as built-up areas of at least 2000 people each without discontinuities of more than 200 metres. As seen in \textbf{(b)}, the overall trend of $\beta^{(n)}$ vs. $n$ remains the same, independently of the precise definition of a city (shaded region): higher quantiles earn disproportionately more in larger cities than in smaller ones. In black, we show the exponents obtained from assuming that the income distribution of each city follows a Gumbel distribution with its given mean and variance, see Eq. \eqref{eq:betan_model} below. The dashed horizontal line and the corresponding filled circles correspond to a null benchmark, where boxes are aggregated at random to create fictitious cities of various sizes.}
    \label{fig:citydefsandscalings}
\end{figure*}

Somewhat surprisingly, the values of $\beta^{(n)}$ for French cities are very similar those reported in \cite{mora2021scaling} for US cities, with the poorest 30-60\% of the population's income actually {\it reducing} as the city population grows, and the rest seeing higher and higher returns on larger populations (\cref{fig:citydefsandscalings}). The bottom 50\% consistently benefit less than the average increase in growth brought about by larger cities, while the top 50\% sees a higher scaling in total income than average.

Thus we have simultaneously two seemingly contradictory results, \textbf{a)} the lack of a strong trend in the variation of local income with local population (Fig. \ref{fig:nullcorrelationplots}), and \textbf{b)} the clear dependence of average city income (and the income of each quantile) on city size.  This shows at the very least that local factors and interactions that have an effect on a person's revenue extend further than the immediate box of side 200m. This suggests that the scale up to which these factors are important are the scale of the city itself, though this simplification excludes factors like travel and migration between cities, and firms encompassing multiple cities. It helps to see however that defining a city as an entirely random combination of points across the country removes this scaling altogether as in \cref{fig:citydefsandscalings}, again granting credence to cities as an economic unit that plays a role in deciding its income distribution.

A lot has been said about scaling laws recently, especially their strong dependence on the definition of what constitutes an agglomeration unit or an urban area \cite{arcaute2015constructing,cottineau2019defining}. These papers show that a large number of reported non-linear scaling exponents switch from super-linear to sub-linear values (or vice-versa) based on the definition of urban areas, and thus cannot be taken seriously. We see these effects as well, especially in the fact that the total city income scales (weakly) super-linearly for three of our used definitions ($\beta^{inc}_{basin}=1.008$, $\beta^{inc}_{uu}=1.004$, and $\beta^{inc}_{200}=1.015$) and sub-linearly for the fourth ($\beta^{inc}_{50}=0.988$). However, the important result we believe to be robust and significant is the upward trend in the scaling exponents for various income deciles which show a minimum range for the basin case at $\beta^{10}_{basin}-\beta^{1}_{basin}=0.06$ and a maximum for the $\rho=200$ per box cutoff case at $\beta^{10}_{200}-\beta^{1}_{200}=0.11$. Given the shaded range in \cref{fig:citydefsandscalings}, the lowest and highest possible differences between the scaling exponents corresponding to different income deciles are similar, at $\Delta\beta_{min}=0.051$ and $\Delta\beta_{max}=0.127$. These differences signify that if we go from an average French city (80-100k people) to Paris (6-12 million people), we will observe a 25-80\% increase in the highest income decile to lowest income decile ratio. We will see in the next two sections that the existence of a scaling law of the type $N^{\beta^{(n)}}$, with $\beta^{(n)}$ close to unity and increasing with decile $n$, can be fully explained by the logarithmic dependence of the mean and standard deviation of income with city size. 

\subsection{Relation between city size and mean/standard deviation of income}

\begin{figure}
     \centering
     \includegraphics[width=\linewidth]{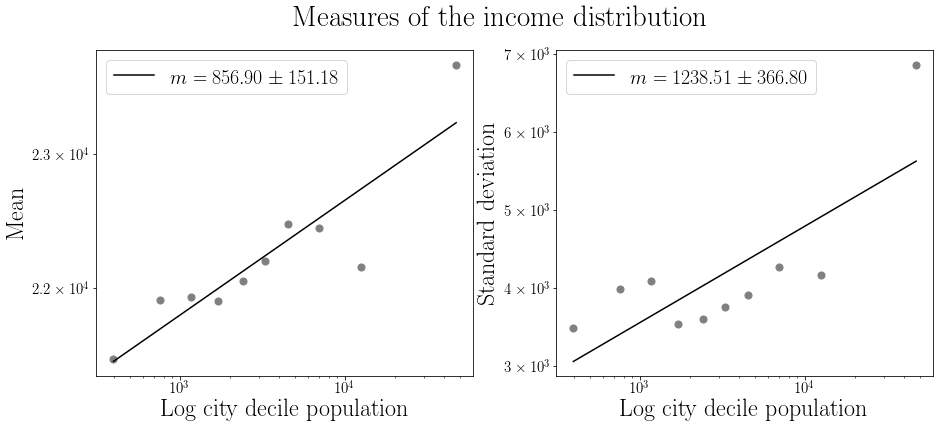}
     \caption{The trends in the average and the standard deviation of the income distribution implies that we have a gradually shifting and widening distribution of revenues for larger and larger cities. Higher order cumulants (skewness, kurtosis) of income do not show a consistent trend as a function of city population. Lines are linear regressions with respect to the logarithm of city size.}
     \label{fig:musigskewkurt}
 \end{figure}

The effect of urban agglomerations on the increase in per-capita income is well-characterized, and we see this in our data in the way the mean income of cities increases with population. Fig. \ref{fig:musigskewkurt} (top left) shows a scatter plot of mean income $\mu$ as a function of population $N$, for different cities, here defined as continuous built-up areas, corresponding to $\beta_{uu}$ in \cref{fig:citydefsandscalings}. We find a statistically significant increase of $\mu$ that can be fitted as $\mu(N) = m_\mu \log_{10} N + c_\mu$ with $m_\mu \approx 860$ and $c_\mu \approx 19200$, both units in euros.

The standard deviation $\sigma$ of incomes exhibits an even stronger dependence on city population, showing that larger cities have huge disparities in income compared to smaller ones. 
 Fig. \ref{fig:musigskewkurt} (top right) shows that $\sigma$ can also be fitted as $\sigma(N) = m_\sigma \log N + c_\sigma$ with $ m_\sigma \approx 1240$ and and $c_\sigma\approx-160$. As we discuss in the next section, the fact that $m_\sigma > m_\mu$ is the basic reason explaining why the scaling exponent $\beta^{(n)}$ is smaller than one for low deciles and larger than one for high deciles. 

\subsection{Universality of the demeaned and rescaled distribution of income}

\begin{figure}[b]
    \centering
    \includegraphics[width=\linewidth]{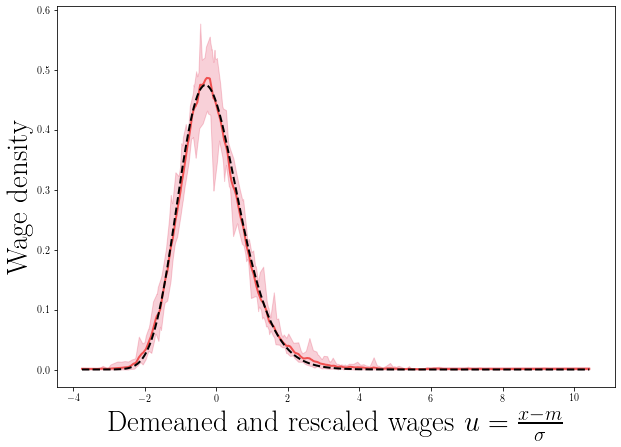}
    \caption{The distributions of demeaned and rescaled income for ten deciles of city sizes, depicted as the red area filled between the min and the max of {\it all} ten individual distributions. The thick red line is the average of these distributions. The dashed black line represents a fit with a GGD (defined in Eq. \eqref{eq:GGD}) with $a=3.90$, $b=0.56$, and $c=2.60$.}
    \label{fig:gumbelfit}
\end{figure}

When extending the analysis of the previous section to the skewness and kurtosis of the income distribution, we find nearly no dependence at all on population size. This suggests that once demeaned by $\mu(N)$ and rescaled by $\sigma(N)$, the distribution of income should become independent of city size. This is confirmed in \cref{fig:gumbelfit}, where we show the distribution of income, demeaned and rescaled, for ten deciles of city sizes. As illustrated by the width of the shaded red area, there is no systematic differences between these empirical distributions (more formally, the distance between this distributions measured as the average pairwise Kolmogorov-Smirnov test statistic scaled with sample size was found to be 0.292, with a standard deviation of 0.120). Averaging over all city sizes, we can fit the resulting universal distribution rather well with a Generalized Gumbel Distribution (GGD), to wit
\begin{equation} \label{eq:GGD}
    \rho(u) = Z(a,b,c) e^{-a e^{-bu} - cu}, \qquad u:= \frac{x - \mu}{\sigma},
\end{equation}
where $x$ is the income, $\mu, \sigma$ the mean income and its standard deviation corresponding to the city size, and $Z$ the normalisation coefficient. \footnote{Explicitly: $Z = \Gamma(c/b) a^{-c/b}/b$.} The coefficients $a,b,c$ are found to be equal to 3.90, 0.56, and 2.60, respectively. 

\begin{table}[h!]
\centering
\begin{tabular}{ |p{1cm}|p{1cm}|p{1cm}|p{1cm}|  }
 \hline
 $n$ & $M_n$ & $n$ & $M_n$ \\ 
 \hline
 1 & -1.474 & 6 & -0.061 \\ 
 2 & -0.990 & 7 & 0.170 \\
 3 & -0.716 & 8 & 0.439 \\
 4 & -0.488 & 9 & 0.800 \\
 5 & -0.275 & 10 & 1.579 \\ 
 \hline
\end{tabular}
\caption{Values of the average of each decile of the Generalized Gumbel Distribution for the parameters we use, i.e. $a=3.90$, $b=0.56$, and $c=2.60$.}
\label{table:1}
\end{table}

Taking such a universal distribution seriously, we can revisit the result of sec. \ref{sec:scaling} on the dependence of the scaling exponent $\beta^{(n)}$ on $n$. First, define the average decile of the GGD as
\begin{equation}
    M_n = \int_{d_{n-1}}^{d_{n}} {\rm d}u \, u \, \rho(u),
\end{equation}
where $d_n$ is such that 
\[ 
\int_{-\infty}^{d_{n}} {\rm d}u \, \rho(u) = n/10
\]
for $n=1,\ldots,9$ and $d_{10}=+\infty$. The numerical values of $M_n$ are given in \cref{table:1}. 
We then define the total income in the $n$-th decile $I_n$ for a city of population $N$ as
\begin{equation}
    I_n = N \left(m(N) + \sigma(N) M_n \right),
\end{equation}
or, using the logarithmic fits of $m(N)$ and $\sigma(N)$ obtained in the previous section, 
\begin{equation}\label{eq:betan_model}
    I_n = (c_\mu + c_\sigma M_n) N \left(1 + K_n \log_{10} N \right), \,\, K_n := \frac{m_\mu + m_\sigma M_n}{c_\mu + c_\sigma M_n}.
\end{equation}
But since $K_n$ is numerically small, we can rewrite $I_n$ as 
$(c_\mu + c_\sigma M_n) N^{\beta^{(n)}}$ with $\beta^{(n)} \equiv 1 + \ell K_n$ with $\ell:=\log_{10} \text{e}$. The corresponding prediction is shown as the thick black line in Fig. \ref{fig:citydefsandscalings}, and is in qualitative agreement with the direct determination of the exponents $\beta^{(n)}$. Note that when the mean and the standard deviation evolve in sync with population size, one finds that $\beta^{(n)}$ is independent of $n$. The difference $\beta^{(n)}-1$ is expected to change sign when $m_\sigma$ is large compared to $m_\mu$.

Such an analysis leads to two important conclusions, not emphasized enough (in our view) in Ref. \cite{mora2021scaling}:  
\textbf{a)} The fact that the difference between the scaling exponents $\beta^{(n)}$ are close to unity reflects an underlying dependence of the mean and standard deviation of the income on the {\it logarithm} of population size; \textbf{b)} The fact that low incomes grow {\it slower} than population size comes from the fact that higher mean income in large cities are not enough to counterbalance stronger income inequalities.

Following up on the last point, we note that if the distribution of income is a universal GGD with mean $m(N)$ and standard deviation $\sigma(N)$, it is not difficult to show that the Gini coefficient $\mathcal{G}$ is given by:
\begin{equation}
    \mathcal{G}(N) = Z'(a,b,c) \frac{\sigma(N)}{m(N)}, 
\end{equation}
where $Z'(a,b,c)$ is a comutable numerical constant. This, together with the logarithmic dependence of $m(N)$ and $\sigma(N)$ on $N$ and the fact that $m_\mu \ll c_\mu$, leads to an approximately logarithmic dependence of the Gini coefficient on $N$, as reported in \cite{cottineau2019defining}.

What are the mechanisms leading to a {\it logarithmic} increase of the mean revenue and its standard deviation with city size, and in particular of a stronger dependence of the latter? It seems to us that this is a crucial question on which future research should focus. We outline a few suggestions in the discussion section below. Before addressing those, however, we want to provide several additional empirical results in the next section.


\section{Some additional empirical findings}

\subsection{A closer look at the relation between local population and income} \label{sec:scaling-bis}

\begin{figure}
    \centering
    \includegraphics[width=\linewidth]{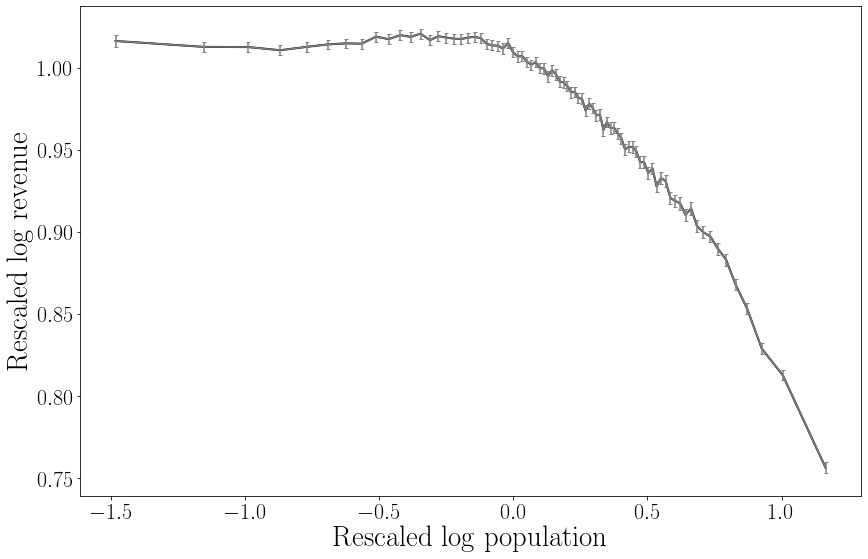}
    \caption{We scale the population of each box with the average box population of the city it belongs to, and its revenue by the average box revenue of the same city. Then we divide the data into $n=81$ quantiles based on this rescaled population, and plot the rescaled revenues of each quantile against it. We get a much better understanding of the underlying information this way than we can from \cref{fig:nullcorrelationplots}, as we see that the more populated neighbourhoods of any city are also relatively poorer.}
    \label{fig:rescaledpoprev}
\end{figure}

Let us revisit the dependence of income on local population density by introducing scaled variables: we first rescale the population of each box by its average over the city the box belongs to. We do the same for revenues: rescale the local box revenue by its average over the city the box belongs to. We then plot the average rescaled income as a function of rescaled population, resulting in \cref{fig:rescaledpoprev}. Whereas \cref{fig:nullcorrelationplots} did not show any correlation between local density and income, the rescaled data paints a much clearer picture. Relative to the city characteristics, heavily populated neighbourhoods are clearly poorer, as could have been anticipated -- a correlation that disappears when all the data is mixed together like in \cref{fig:nullcorrelationplots}, and that is even inverted when we group boxes into cities, since larger cities are wealthier. 

Note that one can also detect a mild decrease of income for low density neighbourhoods, which might come from the fact that, in France, suburbs of large cities are both less populated and relatively poorer. One might this expect such a humped shaped distribution not to hold for US data.  

\subsection{Dependence of other social parameters on city size}

\begin{figure}
    \centering
    \includegraphics[width=\linewidth]{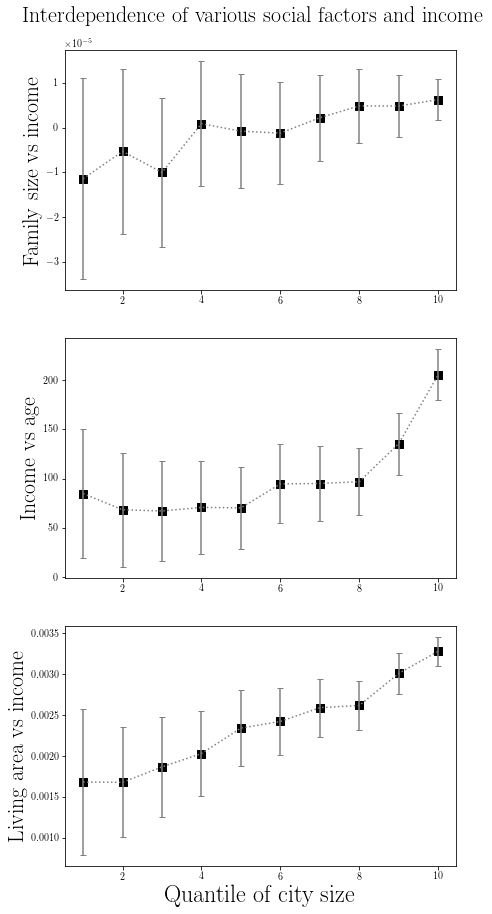}
    \caption{We measure the regressions coefficients $m_i$ of social factors like the average size of a family (top), age (middle), and living surface of families (bottom) with income within each city $i$, and average the slopes of these regressions within each city size decile to get the above plots. These suggest a higher dependence between the variables for larger cities: in the most populated cities, family sizes are larger for higher incomes, income grows strongly with age, and the amount of extra living area you can afford with an increase in salary is higher.}
    \label{fig:socialfactors}
\end{figure}

At this point, we wonder if there are any correlations between the income, city size and other social parameters, like family sizes and age. To this end, we use the data of families per elementary box given to us to calculate the average family size in that box, and we calculate the standard deviation of these average family sizes across each city. Similarly, we have information on how many people in each box correspond to which age bracket. This averaged data can be used to construct a rudimentary age distribution for each city. We look at the correlations between averages and standard deviations of wages, family sizes, and ages. Average income does not seem to show any strong dependences over average family size or average ages, however, the variances of age and family size grow significantly with city population, as does the variance of income.

However, one can study more subtle effects: for each city, we can regress (over different boxes) income vs. age or family size vs. income, etc.. This gives a regression coefficient for each city $i$, call it $m_i$. We then average $m_i$ for all cities of the ``same'' size (i.e. in the same size decile $n$ and plot the result as a function of $n$. These plots exhibit consistent and statistically significant trends (\cref{fig:socialfactors}) that show that: \textbf{(a)} richer families in larger cities are often larger, while richer families in smaller cities seem to be smaller on average, \textbf{(b)} older people have higher incomes across the board, but the age dependency is much stronger in large cities than small towns, and \textbf{(c)} families can afford a larger increase in apartment size when they get a raise in larger cities than in if they get the same raise in income in a smaller city. 

These trends may just be artefacts of nuanced social circumstances: \textbf{(a)} maybe only richer families of those that are large can afford living in big cities; \textbf{(b)} maybe only big cities have corporate chains that go high up enough to provide promotions and raises that increase heavily with age; and \textbf{(c)} perhaps small cities have neither the diverse housing market present in larger cities, in particular nor tiny apartments students and immigrants often live in, explaining the strong impact of income on apartment size. Be that as it may, all of these trends do go hand in hand with the fact that big cities are characterised by a larger dispersion of income.

\subsection{Spatial correlations}

We find that the spatial auto-correlation of population density and of income in each box drops exponentially with the distance between boxes. 
More formally we write
\begin{align} \label{eq:corr_exp}
    \mathbb{E} \left[\frac{(x(r_i)-\mu_x)(x(r_j)-\mu_x)}{\sigma^2_x}\right] \propto \exp\left({-\frac{|r_i-r_j|}{d_c}}\right)
\end{align}
where $x(r_i)$ is value of a certain variable whose covariance we are calculating at position $r_i$, $\mu_x$ and $\sigma_x$ are its mean and standard deviation respectively, and $d_c$ is the characteristic distance associated with the decay of this correlation. We find that the characteristic length of the system is similar whether we look at the correlation of population or of revenue. The length scale $d_c$ over which these spatial correlations decay is $\approx 60$km. We see secondary peaks in the correlation functions at large distances, which is approximately how far large French cities are from each other. In particular, coastal cities in France tend to be more populated and richer, leading to such correlation ``resurgences'' at large distances. 

Such a short-range, exponential decay of correlations is at variance with the observation that cultural traits (such as voting propensity for example) are long range correlated, see e.g. \cite{borghesi2010spatial,borghesi2012election,bouchaud2014emergence}. This is indeed intuitive: whereas opinions, beliefs or preferences can be transmitted from one agent to another by e.g. word of mouth and propagate out to large distances, as argued in \cite{borghesi2010spatial,borghesi2012election,bouchaud2014emergence}, income and population are of fundamentally different nature and do not easily propagate spatially.

\begin{figure}
\hspace{-0.5cm}
    \subfloat[]{
        \includegraphics[width= 0.45\linewidth]{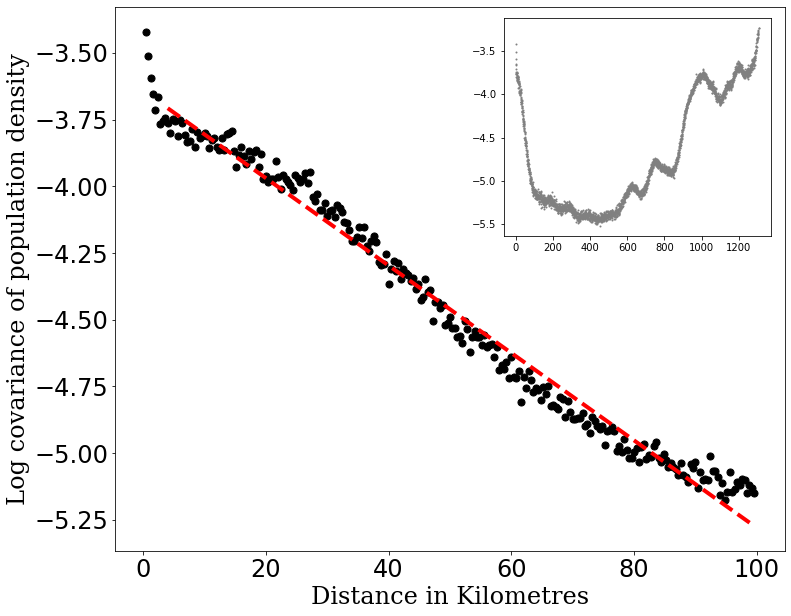}
        }
    \subfloat[]{
        \includegraphics[width= 0.45\linewidth]{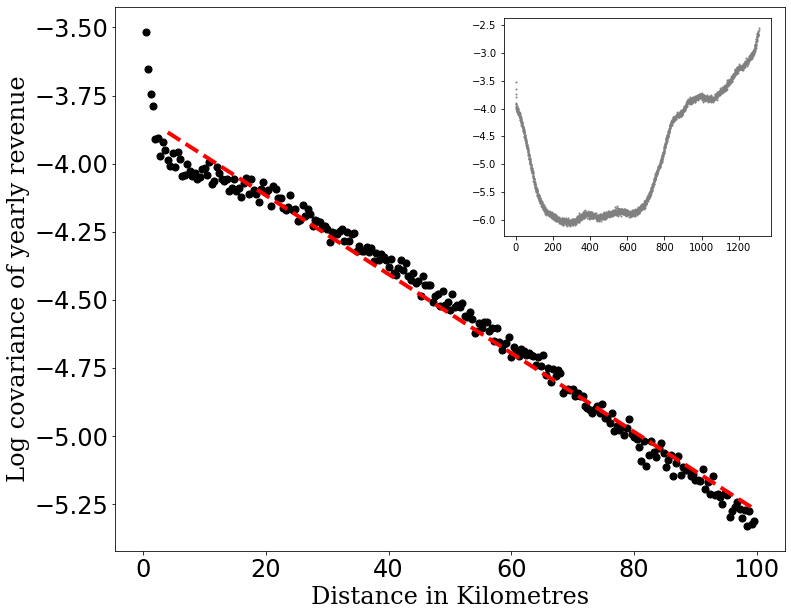}
        }
    \caption{Here we show the spatial auto-correlation of the population density \textbf{(a)} and locally averaged revenues \textbf{(b)}. After a fast decay corresponding to intra-city distances, the decay of correlations in the range 5-100 km is roughly linear in log scale, corresponding to Eq. \eqref{eq:corr_exp} (see dotted red lines). From the slope, we get a characteristic length $d_c$ around 68 km for the population density, and around 60 km for revenue.}
\end{figure}

We have also examined the spatial correlation of income at the smallest distance available, i.e. the standard deviation of income difference between two nearby boxes, measuring the degree of social segregation between close-by neighbourhoods. We have in particular computed the dependence of this standard deviation on the size of the city within which these boxes are embedded. The idea was to test whether rich/poor neighbourhoods are more segregated in large cities, which could have been a mechanism explaining why the average income in each box is more heterogeneous in large cities than in small cities. However, we found that the standard deviation of income difference between neighbouring boxes tracks, up to a constant multiplicative factor, the standard deviation of income of the whole city. This means that larger cities are not more segregated socially than small cities, in agreement with the results shown in \cite{cottineau2019defining}. Hence, the observed increase of inequalities in large cities is not an artefact of the data induced by a stronger spatial segregation.  

\subsection{Scaling of housing costs}

To round off our analysis of inequalities, we investigate the scaling of housing costs across deciles using geo-localized data from the INSEE on house purchases in 2022. The scaling exponents for each house-price decile are found to be significanlty larger than those for income scaling, indicating a steeper increase in housing costs with city population. While the trend between the exponents of different deciles for housing costs is not clear, it suggests that the most and the least expensive parts of the house-price spectrum see the largest increase as a function of city population. 

\begin{figure}[t]
    \centering
    \hspace{-1.5cm}
    \includegraphics[width=\linewidth]{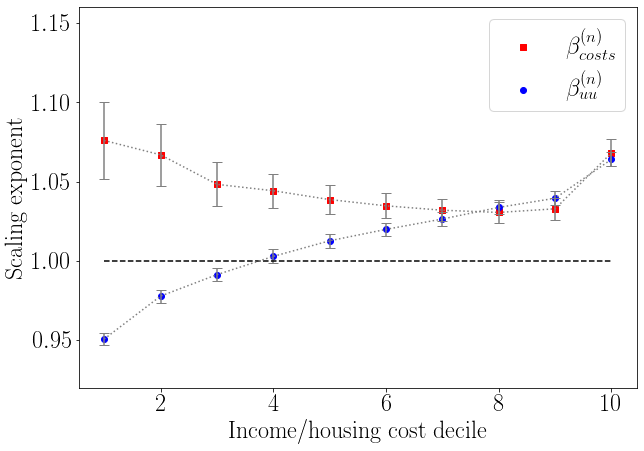}
    \caption{Comparison between scaling exponents for housing costs versus those for income for the urban units case.}
    \label{fig:costsexp}
\end{figure}

It is worth noting that the data only represents average transaction prices of houses, which we use as a proxy for living costs due to a lack of average rent data. It is uncertain how similar the trends in apartment prices and rents are, though we may assume rents also scale super-linearly. As mortgages and rent for the same houses are typically similar \cite{lo2022causal,gallin2008long}, we can reason that rent scales at higher rates than prices, as the profit margin, given by the difference between rent and mortgage, should have scaling exponents similar to that of income. This is because a significant portion of the income of the richer deciles likely comes from rent (considering for example the revenue split of one of the richer parts of Paris \cite{rentproof75106}). This implies that the scaling for percentage income spending on housing will see a sharp negative trend throughout deciles, if we consider it to be at the very least equal to the difference between the cost and income exponents we observe in \cref{fig:costsexp}. Again, as the richer deciles do not pay rent, this will only be a good approximation till decile 8 or 9, and the sudden rise in house cost exponent for the most expensive houses will not feature into their living costs as it would for the poorest. 

Overall this negative trend over deciles for costs in addition to the positive trend for incomes suggests that the effective inequality, for example the dispersion of disposable income available after living costs, is actually much larger than the dispersion of pre-cost incomes.

\section{Discussion}

\subsection{Qualitative considerations} 

The reasons behind the sub-linear/super-linear divide of income scaling of different deciles could have to do either with \textbf{(a)} an absolute scaling in average income of a particular job based on city population, or \textbf{(b)} a scaling of the compositions of the salaried workers population of each city (or both). The former would imply the existence of jobs which are paid worse (or better) as city population grows. This is hard to imagine, but might be true if we factor in increased competition or discrimination against immigrants in larger cities for lower-skilled jobs, resulting in a drop in wages \cite{weichselbaumer2007effects}. It could also be due to the dependence of availability and wages of certain jobs on other factors that are known to scale with city population. Perhaps low paying jobs scale with a measure of city infrastructure per capita, something known to scale sub-linearly with population.
\begin{figure}
    \centering
    \includegraphics[width=\linewidth]{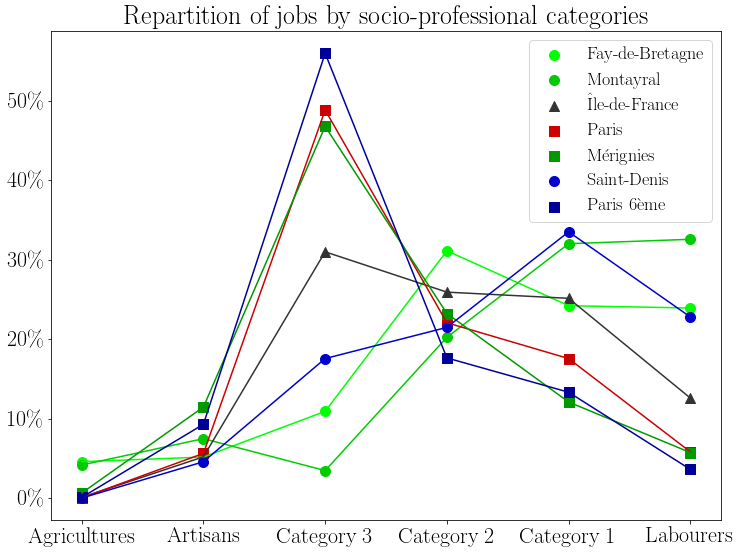}
    \caption{The partition of jobs by socio-professional categories as defined in the INSEE databse in randomly chosen postal codes in small cities, and the richest and poorest parts of Paris. We see that the distribution of jobs within these postal codes depends less on how populated the region is and more on whether its population is rich (or rather vice-versa). The affluent commune of Mérignes has the same kind of job distribution as the 6th arrondisement of Paris, while people working in Fay-de-Bretagne and  Montayral comprise of similar socio-professional categories as those in Saint-Denis. These distributions are found to be generic, with all French communes displaying a job profile lying between the two extremes. We also show the distributions for all of Paris in red which naturally includes its 6th district, and the entire Île-de-France region in grey which contains both Paris and the suburb of Saint-Denis in its confines. Looking at larger scales this way has an effect of averaging out the socio-professional categories. Thus it is clear that different incomes in different cities is not just the effect of an absolute change in pay for the same jobs. Categories are defined in \cite{categories}. }
    \label{fig:jobdist}
\end{figure}

Similarly, maybe wages in high paying jobs scale with the number of human interactions, which clearly follows a super-linear scaling with respect to city population. The hierarchical structure of big firms, where each person manages $k$ people who, in turn, manage $k$ persons each, may also lead to a non-trivial scaling of a person's income with the size of the firm \cite{montroll19821}. On the other hand, the sub-linear/super-linear divide may also be due to entirely different jobs that are more prevalent in large cities, for example childcare and retail on the lower end of the spectrum, and stock traders and advertising executives in higher quantiles. It could also be due to the disappearance of land intensive occupations like people who own wine-yards or farms. This last example could point to the fact that both possible reasons go hand in hand, as the available land for such occupations decreasing would first lead to a decrease in income of such jobs and then the disappearance of the occupations altogether. 

We check whether there is some truth to this assumption by looking at the exact kind of jobs people do in different parts of France, shown in \cref{fig:jobdist}. We see that there is a clear difference between kind of jobs prevalent in rich and poor neighbourhoods. The categories correspond to the INSEE definitions, with ``Category 3" corresponding to the group \textit{Cadres et professions intellectuelles supérieures}, typically high paying jobs, ``Category 1" being the group \textit{Employés} which are usually jobs that don't pay a lot, while ``Category 2" corresponds to \textit{Professions intermédiaires} which lies somewhere between the other two categories. \textit{Labourers} are perhaps the least paid of all socio-professional groups, while \textit{Agriculturists} and \textit{Artisans} can cover a large range of revenues (the latter includes shop owners, hair-stylists, as well as owners of big businesses) and thus are difficult to rank in this manner. Further definitions on what comprises these groups is given in \cite{categories}. What we gather from this visualisation is that the differences in income in various regions cannot be solely due to the same jobs paying different amounts, and that different compositions of the job market are clearly an important factor.

\begin{figure}
    \centering
    \hspace{-1cm}
   \vspace{-0.5cm} \includegraphics[width=\linewidth]{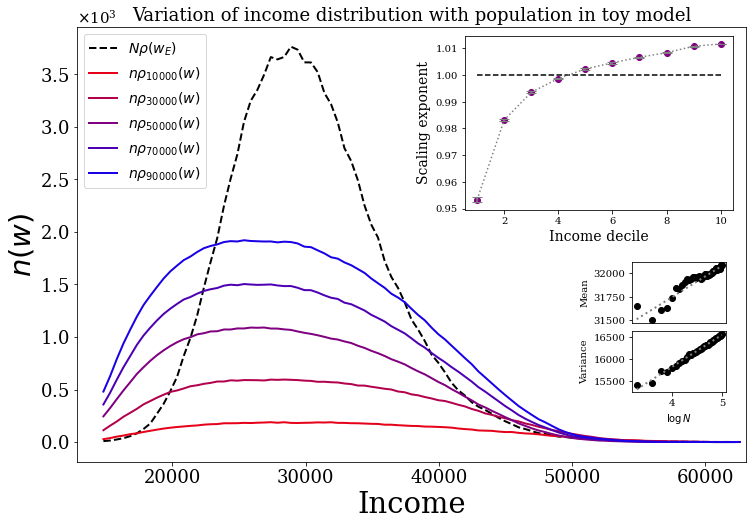}
    \caption{In the simulations of our toy model described by \cref{eq:toymodelprobs}, we get an income distribution with an increasing mean and standard deviation as we grow the size of the city. Both growths are linear, with the latter being much faster than the former, leading to the same kind of city size effects as those found in the French data. For the simulation resulting in these distributions we have used $c_E=0.5$, $c_\rho=0.001$, and $\gamma=3$. The mean and standard deviation of the distribution are found to increase logarithmically with the population, with $\mu_w \approx 300 \log_{10}N +c_\mu$, and $\sigma_w \approx 750 \log_{10}N+c_\sigma$, thus qualitatively displaying a similar behaviour as observed due to the values in \cref{fig:musigskewkurt}.}
    \label{fig:toymodeldists}
\end{figure}

\subsection{A toy-model}
To quantify the effects of crowding in cities, we construct a simplified toy-model. Imagine that people have expected incomes -  based on their circumstances, education, luck, skill, etc - that are spread according a Gumbell distribution of the kind we found in the previous sections. We consider the following dynamics: an initially empty city  progressively grows into a city of size $N$, which can be assumed to be via birth or migration. We write the probability of getting a job of income $w_{n+1}$ when joining a city which already has size $n$ as
\begin{align}
\begin{aligned}
    P_J(w_{n+1}\vert w_E) &\propto \exp\left[-\left\vert\frac{w_E-w_{n+1}}{w_Ec_E}\right \vert ^{\gamma}\right. \\
    &\left.-\log (1+n)\left(\frac{\rho_n(w_{n+1})\Delta w}{c_\rho}\right)\right],\label{eq:toymodelprobs}
    \end{aligned}
\end{align}
where $w_E$ is the expected income for each person and $\Delta w$ is the width of the income brackets that we consider, making $\rho(w)\Delta w$ the fraction of people in said income brackets. Thus brackets that are already densely occupied are less accommodating to new arrivals, pushing people to the edges. The effect of which is that a larger population is 'creating' jobs with higher and lower incomes. At the same time, people are reluctant to take jobs that pay less than what they reasonably expect to be paid due to their circumstances and education. On the other hand, jobs that pay more than the average pay at a person's level of skills or contacts are harder to get as well, all the more so in large cities (hence the factor $\log(1+n))$.\footnote{While we use a single exponent $\gamma$ for this effect, it can instead be modelled as two different exponents to quantify the asymmetry in getting jobs that pay more or less than what a person wants or is trained for. }

In this simple simulation, we do not modify previous people's incomes as the city grows, neither do we let them find different jobs, or adjust their expected income with time in the form of further acquired skills, capital, or promotions. All these changes would help fit our toy model more perfectly to the data we analysed in this article. As we do however obtain behaviour similar to what we want, we argue that competition could be a major driving factor in the larger spread of incomes in larger cities. 

\section{Conclusion} 

Previous works suggest that quantities that scale super-linearly are those depending on interactions within the population, whereas quantities that scale sub-linearly depend on city infrastructure, which evidently often does not keep up with city population \cite{barthelemy2019statistical}. As we find that the same measure, i.e. total income, displays either regime of scaling based on what part of the population we are looking at, salaries of different kind of jobs might also be dependent on fundamentally different factors for different parts of the population. In the face of valid criticism for scaling laws \cite{leitao2016scaling}, we make various measurements on our data to show that inequalities are without doubt increasing with city population. We were able to reach a number of interesting qualitative and quantitative conclusions: \textbf{(a)} the rise in inequalities as a function of city size regardless of city definition; \textbf{(b)} the larger dispersion of social factors like family sizes and living areas in bigger cities; \textbf{(c)} the universality of rescaled income distribution and finally \textbf{(d)} the logarithmic dependence of the mean and standard deviation of income as a function of city size, that (together with 
\textbf{(c)}) naturally explains the different scaling exponents for different income deciles.

While it is often suggested that urban areas lead to a higher productivity due to being ``agglomeration economies" \cite{rosenthal2004evidence}, we find that these increasing returns are skewed in favour of the rich. The median income of the population sees very little increase with city size and the mode actually shifts downwards, which calls into question the supposed benefits of agglomeration economies. Cities are not closed systems, and research suggests migration can dominate the population dynamics of cities \cite{verbavatz2020growth}, however there is evidence from elsewhere in the world that suggests that travel for the purpose of earning one's living is most often short term, especially among the poor \cite{banerjee2007economic}, meaning that this population is probably not well captured by official statistics. Hence, given the insurmountable evidence for it in today's world as both a source of increased poverty and improved living standards \cite{czaika2012role,beegle2011migration}, migration is most likely an important factor but there are probably other reasons unrelated to migration causing the larger dispersion of wealth in urban areas.

With the higher costs of upkeep/lower sustainability of urban regions mixed with higher inequalities, rising emissions, higher temperatures, and risk of global pandemics, and waves of immigration, we might be at the dawn of an age that requires de-urbanisation. The transition to an equitable society will have to start with policies directed at balancing these challenges, for greater social welfare and for reducing unrest as our challenges and economic inequalities increase.


\section*{Acknowledgements}

We would like to thank Marc Barth\'el\'emy for his insightful comments on scaling laws, along with Fabián Aguirre López for pointing us to the relevant literature. We also acknowledge discussions with Antoine-Cyrus Becharat on the issue of housing prices, Jerome Garnier-Brun for his help with the plots and their interpretations, Cecilia Auburn, Natascha Hey, and Anirudh Pammi for sharing their wisdom in numerical analysis, and Swann Chelly for pointing us towards the data we needed for this study.

\bibliography{apssamp}

\end{document}